# Ultra-low magnetic damping of a metallic ferromagnet

Martin A. W. Schoen,[1, 2] Danny Thonig,[3] Michael L. Schneider,[1] T. J. Silva,[1] Hans T. Nembach,[1] Olle Eriksson,[3] Olof Karis,[3] and Justin M. Shaw[1*]

[1]Quantum Electromagnetics Division, National Institute of Standards and Technology, Boulder, CO, 80305, USA
[2]Institute of Experimental and Applied Physics, University of Regensburg, 93053 Regensburg, Germany
[3]Department of Physics and Astronomy, University Uppsala S-75120 Uppsala, Sweden

*Corresponding author: justin.shaw@nist.gov

**The phenomenology of magnetic damping is of critical importance for devices that seek to exploit the electronic spin degree of freedom since damping strongly affects the energy required and speed at which a device can operate. However, theory has struggled to quantitatively predict the damping, even in common ferromagnetic materials[1–3]. This presents a challenge for a broad range of applications in spintronics[4] and spin-orbitronics that depend on materials and structures with ultra-low damping. Such systems enable many experimental investigations that further our theoretical understanding of numerous magnetic phenomena such as damping and spin-transport[5] mediated by chirality[6] and the Rashba effect. Despite this requirement, it is believed that achieving ultra-low damping in metallic ferromagnets is limited due to the scattering of magnons by the conduction electrons. However, we report on a binary alloy of Co and Fe that overcomes this obstacle and exhibits a damping parameter approaching $10^{-4}$, which is comparable to values reported only for ferrimagnetic insulators[7,8]. We explain this phenomenon by a unique feature of the bandstructure in this system: The density of states exhibits a sharp minimum at the Fermi level at the same alloy concentration at which the minimum in the magnetic damping is found. This discovery provides both a significant fundamental understanding of damping mechanisms as well as a test of the theoretical predictions put forth by Mankovsky et al.[3].**

In recent decades, several theoretical approaches have attempted to quantitatively predict magnetic damping in metallic systems. One of the early promising theories was that of Kambersky, who introduced the so-called breathing Fermi surface model.[9–11] More recently, Gilmore and Stiles[2] as well as Thonig et al.[12] demonstrated a generalized torque correlation model that include both intraband (conductivity-like) and interband (resistivity-like) transitions. The use of scattering theory to describe damping was later applied by Brataas et al.[13] and Liu et al.[14] to describe damping in transition metals. A numerical realization of a linear response damping model was implemented by Mankovsky[3] for Ni-Co, Ni-Fe, Fe-V and Co-Fe alloys. For the Co-Fe alloy, these calculations predict a minimum intrinsic damping of $\alpha_{int} \approx 0.0005$ at a Co-concentration of 10 % to 20 %, but was not experimentally observed.[15]

Underlying this theoretical work is the goal of achieving new systems with ultra-low damping that are required in many magnonic and spin-orbitronics applications.[7,8] Ferrimagnetic insulators such as yttrium-iron-garnett (YIG) have long been the workhorse for these investigations since YIG films as thin as 25 nm have experimental damping parameters as low as $0.9\times10^{-4}$.[16] Such ultra-low damping can be achieved in insulating ferrimagnets in part due to the absence of conduction electrons and therefore the suppression of magnon-electron scattering. However, insulators cannot be used in most spintronic and spin-orbitronic applications where a charge current through the magnetic material is required nor is the requirement of growth on gadolinium gallium garnet templates compatible with spintronics and complementary metal-oxide semiconductor (CMOS) fabrication processes. One proposed alternative class of materials are Heusler alloys, some of which are theoretically predicted to have damping parameters as low as $10^{-4}$.[17] While such values have yet to be realized, damping parameters as low as 0.001 have been reported for $Co_2FeAl$[18] and NiMnSb[19]. However, Heusler alloys have non-trivial fabrication constraints, such as high-temperature annealing, which are also incompatible with spintronic and CMOS device fabrication constraints.

In contrast, metallic ferromagnets such as 3d transition metals are ideal candidate materials for these applications since high quality materials can be produced at room temperature (RT) without the requirement of annealing. However, ultra-low damping is thought to be unachievable in metallic systems since damping in conductors is dominated by magnon-electron scattering in the conduction band resulting in a damping parameter over an order of magnitude higher than those found in high-quality YIG.

Inspired by Mankovsky's theoretical prediction of ultra-low damping in the $Co_xFe_{1-x}$ alloy system[3], we systematically studied the compositional dependence of the damping parameter in $Co_xFe_{1-x}$ alloys, including careful evaluation of spin-pumping and radiative damping contributions. Polycrystalline $Co_xFe_{1-x}$ alloy films, 10 nm thick, were sputter-deposited at RT with Cu/Ta seed and capping layers. X-ray diffraction (XRD) reveals that the CoFe alloys display a body-centered-cubic (bcc) phase over a Co concentration of 0 % to 60 %, a face-centered-cubic (fcc) phase above 80 % Co, and a mixed phase between 60 % and 80 % Co, in good agreement with the bulk phase diagram of this system. The damping parameter is determined from broadband ferromagnetic resonance (FMR) spectroscopy which measures the susceptibility over frequencies spanning 5 GHz to 40 GHz. An example of $S_{21}(H)$ vector-network-analyzer transmission data is shown in Figure 1a and b, together with fits to the complex susceptibility for the real and imaginary parts, respectively. The total damping parameter $α_{tot}$ is determined from the frequency dependence of the linewidth obtained from these susceptibility fits, according to equation (1),

$$\Delta H = \frac{2h\alpha_{tot}}{g\mu_0\mu_B} f + \Delta H_0 \qquad (1)$$

where $\mu_0$ is the vacuum permeability, $\mu_B$ is the Bohr magneton, $h$ is Planck's constant, $g$ is the Landé $g$-factor, and $\Delta H_0$ is the inhomogeneous linewidth.

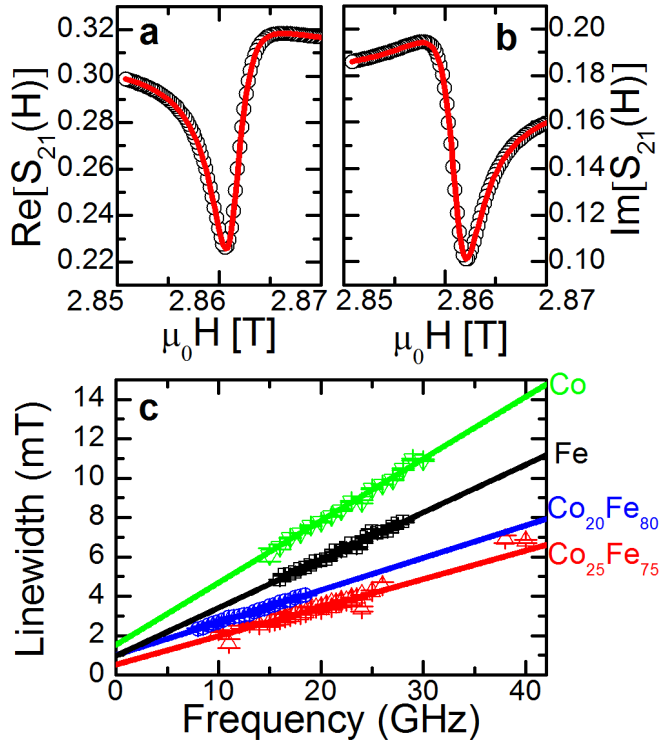

**Figure 1: Ferromagnetic resonance spectra, measured via FMR and the resulting linewidth as a function of frequency. a** and **b**, respectively, show the real and imaginary parts of the $S_{21}(H)$ transmission parameter (black circles) with the complex susceptibility fit (red lines). **c,** the line-widths (symbols) are plotted versus the frequency for Co, Fe, $Co_{20}Fe_{80}$ and $Co_{25}Fe_{75}$. The uncertainties in the line-widths were obtained by means of the standard method for the determination of confidence limits on estimated parameters for nonlinear models under the assumption of Gaussian white noise. The lines are error-weighted fits to equation (1), which are used to determine both the total damping $\alpha_{tot}$ and the inhomogeneous linewidth broadening for each alloy.

The measured total damping $\alpha_{tot}$ versus alloy composition for 10 nm films is plotted in Figure 2. $\alpha_{tot}$ shows a clear minimum of $(2.1 \pm 0.15) \times 10^{-3}$ at a Co concentration of 25 %. However, as a result of the utilized measurement geometry and the structure of the sample, there are several extrinsic contributions to $\alpha_{tot}$ that are independent of $\alpha_{int}$.

The first contribution—the result of the inductive coupling of the precessing magnetization and the coplanar waveguide (CPW)—is radiative damping $\alpha_{rad}$.[20] The FMR system is designed and optimized to couple microwaves into the ferromagnet, and therefore, by virtue of reciprocity, the system is efficient at coupling microwaves out of the ferromagnet. For very thin films or films with low saturation magnetization, $\alpha_{rad}$ is typically not a significant contribution to the total damping and can be ignored. However, in the present case, $\alpha_{rad}$ must be accounted for in the analysis due to the combination of a very high saturation magnetization and the exceptionally small value of $\alpha_{int}$. As described in the supplemental information section (SI), we calculate and experimentally validate the contribution of $\alpha_{rad}$ to the total damping, which is plotted in Figure 2.

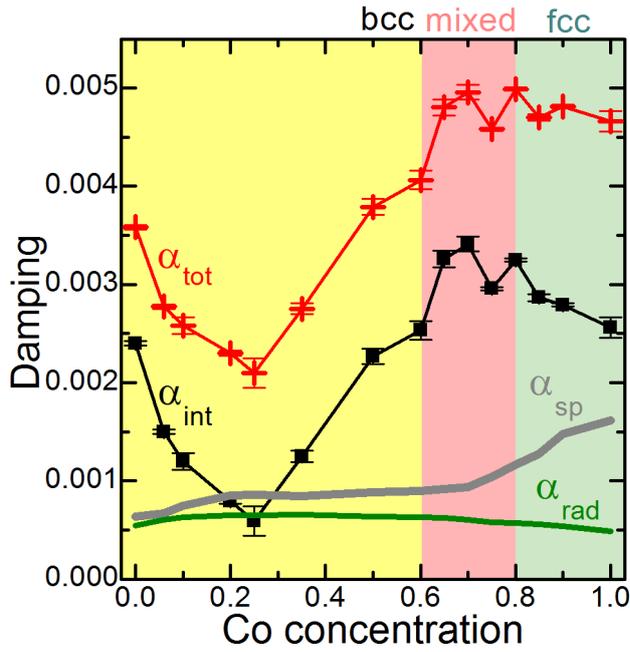

**Figure 2: The total measured damping with radiative and interfacial contributions.** The total damping $\alpha_{tot}$ (red crosses with lines), spin- pumping $\alpha_{sp}$ (gray line) and radiative $\alpha_{rad}$ (green line) contributions to the damping, and the intrinsic damping $\alpha_{int}$ are plotted against the respective Co concentration. The errors are propagated from the line-width fits. The crystal structure of the alloys, obtained from XRD, signified by the color regions in the plot.

The second non-negligible contribution to the total damping is the damping enhancement due to spin-pumping into the adjacent Cu/Ta layers. The spin-pumping contribution $\alpha_{sp}$ can be determined from the thickness dependence of ($\alpha_{tot}$ - $\alpha_{rad}$) since it behaves as an interfacial damping term[21]. Indeed, we measured the thickness dependence of ($\alpha_{tot}$ - $\alpha_{rad}$) for many alloy samples in order to quantify and account for $\alpha_{sp}$ (see SI), which is displayed in Figure 2.

Contributions from eddy-current damping[22] are estimated to be smaller than 5% and are neglected. Finally, two-magnon scattering is minimized in the perpendicular geometry used in this investigation and its contribution is disregarded[23].

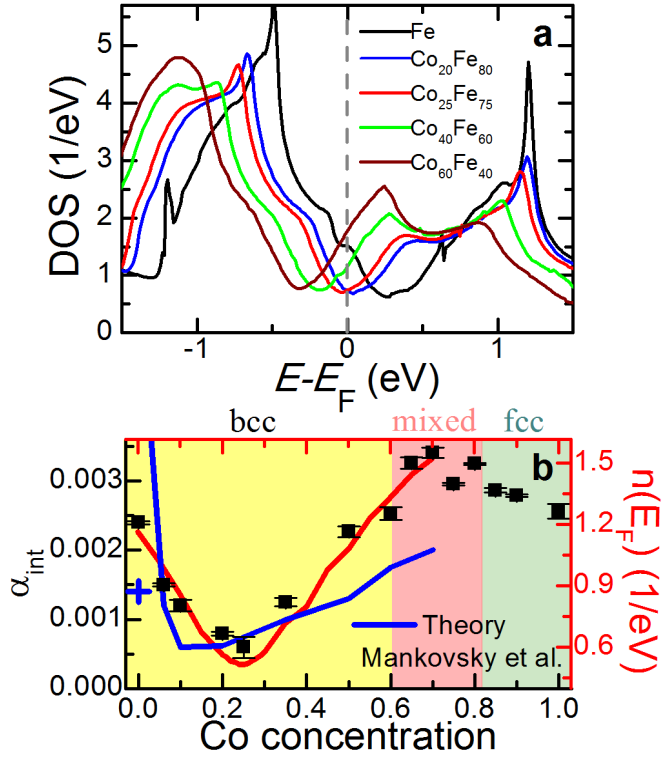

**Figure 3: Calculated electron density of states (DOS) and its comparison to the intrinsic damping.**
**a** Electronic structure of bulk $Co_xFe_{1-x}$. The DOS is shown for different Co concentrations, as indicated. $E_F$ is the Fermi energy. **b** The extracted intrinsic damping (black squares, left axis) is compared to the theory in Mankovsky et al.[3], for a temperature of 0 K (blue line) and for a temperature of 300 K for pure Fe (blue cross). The calculated DOS at the Fermi energy $n(E_F)$ is plotted on the right axis (red line). The y-offset of $n(E_F)$ is chosen deliberately in order to demonstrate that the concentration-dependent features of the damping directly correlate to features of $n(E_F)$. We cannot exclude concentration-independent contributions to the damping, which are accounted for by the 0.4 eV$^{-1}$ y-offset.

The total measured damping becomes, $\alpha_{tot} = \alpha_{int} + (\alpha_{rad}/2) + \alpha_{sp}$, allowing the intrinsic damping $\alpha_{int}$ to be determined, which is presented in Figure 2. For many values of $\alpha_{int}$, the contributions of $\alpha_{sp}$ and $\alpha_{rad}$ are of similar magnitude, showing the importance of accounting for these contributions. For 25 % Co $\alpha_{int}$ now displays a sharp minimum in damping of $(5 \pm 1.5) \times 10^{-4}$, which is astonishing for a conductor. Indeed, $\alpha_{int} < 0.001$ have been measured only in ferrimagnetic insulators.[24]

These results raise the question why $\alpha_{int}$ can be so low in the presence of conduction electrons. To gain a deeper understanding, we performed electronic structure calculations for $Co_xFe_{1-x}$ within a full-relativistic, multiple-scattering approach (Korringa-Kohn-Rostoker method[25], KKR) using the coherent potential approximation (CPA)[26,27] over the entire range of compositions (see SI). Several representative examples are given in Figure 3a.

The d−states (peak in the DOS below $E_F$) for pure Fe are not fully occupied. Consistent with the rigid band model[28], the d−states shift to lower energies when the concentration of Co increases, and become fully occupied at 25 % Co, coinciding with the minimum in $n(E_F)$ displayed in Figure 3a, which originates from the hybridization between majority Fe $e_g$ and minority Co $t_{2g}$ states.

Ebert et al.[1] suggested that $\alpha_{int}$ is proportional to $n(E_F)$ in the breathing Fermi-surface model (i.e., intraband transitions) in the cases of a minimally varying spin-orbit coupling (SOC) (as is the case for the $Co_xFe_{1-x}$ system) and small electron-phonon coupling[2,29]. Alternatively, interband transitions become significant only if bands have a finite overlap due to band broadening, caused for example by coupling to the phonons. As a result, interband transitions suppressed at low-temperature and energy dissipation becomes dominated by intraband transitions. Our RT measurements of $Co_xFe_{1-x}$ satisfy this "low-temperature" condition since the electron-phonon coupling is < 20 meV for pure bcc Fe, and < 30 meV for pure hcp Co. Band broadening due to disorder is about 15 meV for the, at $E_F$ dominating, $e_g$ states (50 meV for the $t_{2g}$ states) in $Co_{25}Fe_{75}$ and varies up to 55 meV for the $e_g$ states (150 meV for the $t_{2g}$ states) over the whole range of composition. These calculations show that the band broadening effect at RT is too small to provide significant interband damping, consistent with the almost perfect proportionality between $n(E_F)$ for all alloy compositions in the bcc phase (0 % to 60 % Co). Such a proportionality requires an offset of 0.4 eV$^{-1}$, which originates from the fact the $n(E_F)$ is a superposition of all states, some of which do not contribute significantly to the damping.

The calculations of $\alpha_{int}$ by Mankovsky[3] (included in Fig 3b) show a minimum value of $\alpha_{int} \approx 0.0005$ between 10 % and 20 % Co, which differs from the sharp minimum we find at 25 % Co in both the experimental data and the calculated values of $n(E_F)$. Remarkably, with the exception of pure Fe, the calculated values of $\alpha_{int}$ (at 0 K) agree with our results within a factor of $\approx 2$. Furthermore, the agreement is greatly improved for pure Fe, when a finite temperature of 300 K is included. While not perfect, the agreement between those calculations and our results is compelling and provides the critical feedback needed for further refinement of theory.

We therefore demonstrate and conclude that $\alpha_{int}$ is largely determined by $n(E_F)$ in the limit of intraband scattering. Secondly, our work shows that a theoretical understanding of damping requires an accurate account of all contributions to the damping parameter. Furthermore, if the theoretical explanation put forth here to explain low damping holds in general, it is natural to utilize data-mining algorithms to screen larger groups of materials in order to identify additional low-damping systems. Examples of such studies to identify new materials for use, for example, in scintillators have been published[30], and the generalization to applications in magnetization dynamics is straightforward.

## Author responsibility

MAWS and DT wrote the manuscript, JS and HN conceived of the experiment, MAWS deposited the samples, and performed the SQUID measurements and analysis, MAWS, MLS and HN performed the FMR measurements and analysis, JS performed the XRD measurements and analysis, DT and OE performed the first-principles DFT calculations. All authors contributed to the interpretation of the results.

## Acknowledgements

The authors are grateful to Mark Stiles and Eric Edwards for valuable discussions.

## References.

# Methods: Ultra-low magnetic damping of a metallic ferromagnet

Martin A. W. Schoen,[1,2] Danny Thonig,[3] Michael Schneider,[1] Thomas J. Silva,[1] Hans T. Nembach,[1] Olle Eriksson,[3] Olof Karis,[3] and Justin M. Shaw[1]*

[1]Quantum Electromagnetics Division, National Institute of Standards and Technology, Boulder, CO, 80305
[2]Institute of Experimental and Applied Physics, University of Regensburg, 93053 Regensburg, Germany
[3]Department of Physics and Astronomy, University Uppsala S-75120 Uppsala, Sweden

Dated: 12/11/2015

PACS numbers: 71.70.Ej,76.50.+g, 75.78.n,71.20.Be

*Corresponding author: justin.shaw@nist.gov

**Sample preparation**

The samples were deposited by DC magnetron sputtering at an Ar pressure of approximately 0.67 Pa (5 × $10^{-3}$ Torr) in a chamber with a base-pressure of less than 5 × $10^{-6}$ Pa (4 × $10^{-8}$ Torr). The alloys were deposited by co-sputtering from two targets with the deposition rates calibrated by X-ray reflectometry (XRR). The repeatability of the deposition rates was found to be better than 3% variation over the course of this study. For all deposited alloys, the combined deposition rate was kept at approximately 0.25 nm/s, to ensure similar growth conditions. Furthermore the $Co_{20}Fe_{80}$ and the $Co_{25}Fe_{75}$ samples were replicated by depositing from single stoichiometric targets, to prove the reproducibility of the results. Samples with a thickness of 10 nm were fabricated over the full alloy composition range and additional thickness series (7 nm, 4nm, 3nm and 2 nm) were fabricated for the pure elements and select intermediate alloy concentrations (20% Co, 25% Co, 50% Co and 85% Co).

**X-ray diffraction measurement.**

The crystal structure of the alloys was determined by X-ray diffraction XRD using an in-plane geometry with parallel beam optics and a Cu $K_\alpha$ X-ray source. The in-plane geometry allows the signal from the Co-Fe alloys to be isolated from the high-intensity signal coming from the silicon substrate. These measurements yield both the in-plane lattice constants and the crystal structure, as shown in the supplemental material, section 1. The deposition rates were calibrated using XRR using the same system configured for the out-of-plane geometry.

**SQUID measurement.**

We measured the in-plane hysteresis curves at 300 K to determine the magnetic moment of the sample. Sample were first diced with a precision diamond saw such that an accurate value of the volume of the sample could be calculated. The saturation magnetization $M_S$ for all alloy samples is then determined by normalizing the measured moment to the volume of the CoFe in the sample. These values are shown in the supplemental material.

**VNA-FMR measurement**

The FMR measurement utilized a room-temperature-bore superconducting magnet, capable of applying static magnetic fields as high as 3 T. An approximately 150 nm poly(methyl methacrylate) (PMMA) coat was first applied to the samples to prevent electrical shorting of the co-planar waveguide (CPW) and to protect the sample surface. Sample were placed face down on the CPW and microwave fields were applied in the plane of the sample, at a frequencies that ranged from 10 GHz to 40 GHz. A vector-network-analyzer (VNA) was connected to both sides of the CPW and the complex $S_{21}(H)$ transmission parameter was determined. The iterative susceptibility fit of $S_{21}(H)$ was done with the method described by Nembach et al.[1]. All fits were constrained to a field window that was 3 times the linewidth around the resonance field in order to minimize the fit residual. We verified that this does not change the results, but reduces the influence of measurement noise on the error bars of the fitted values.

**Calculation of the DOS**

Electronic ground state calculations have been performed by a full-relativistic multiple-scattering Green's function method (Korringa-Kohn-Rostoker method[2], KKR) that relies on the local spin-density approximation (LDA) to density functional theory. We utilized Perdew–Wang exchange correlation functionals[3–6].

In our multiple-scattering theory, the electronic structure is described by scattering path operators $\tau_{ij}$ (i,j lattice site indices) [2], where, in the spin-angular-momentum representation, we consider angular momenta up to $l_{max} = 3$ and up to 60 x 60 x 60 points in reciprocal space. The substitutional alloys are treated within the coherent potential approximation (CPA)[7,8]. Co impurities in the Fe host lattice are created in the effective CPA medium by defect matrices. The CPA medium is described by scattering path operators. The site-dependent potentials are considered in the atomic sphere approximation (potentials are spherically symmetric within muffin-tin spheres and constant in the interstitials).

The DOS is obtained from the integrated spectral density[2]

$$n(E) = \frac{1}{\pi} \int_{\Omega_{BZ}} Im[\text{tr}(G(E + i\eta, \boldsymbol{k}))] d\boldsymbol{k}$$

with the small positive energy $\eta$. The integration in reciprocal space $\boldsymbol{k}$ runs over the first Brillouin zone $\Omega_{BZ}$.

# Supplemental Information: Ultra-low magnetic damping of a metallic ferromagnet

Martin A. W. Schoen,[1,2] Danny Thonig,[3] Michael Schneider,[1] Thomas J. Silva,[1] Hans T. Nembach,[1] Olle Eriksson,[3] Olof Karis,[3] and Justin M. Shaw[1]

[1]Quantum Electromagnetics Division, National Institute of Standards and Technology, Boulder, CO, 80305
[2]Institute of Experimental and Applied Physics, University of Regensburg, 93053 Regensburg, Germany*
[3]Department of Physics and Astronomy, University Uppsala S-75120 Uppsala, Sweden

Dated: 12/11/2015

**X-ray diffraction measurement**

Figure S1 shows several in-plane geometry X-ray diffraction (XRD) spectra for a series of select alloys. From 100 % to 80 % Co, there is no evidence of a body-centered cubic peak and we can conclude that the phase of the CoFe is purely face-centered cubic (fcc). At a concentration of 75% (not shown), a bcc peak emerges in the spectrum. An fcc (220) peak is always visible in the spectra due to the signal from the Cu seed and capping layers. The presence of the signal from the Cu layers complicates the analysis since the point at which the crystalline phase transitions from a mixed fcc and bcc to a pure bcc phase for the $Co_xFe_{1-x}$ alloys cannot be exactly determined. However, there is strong evidence that the phase transition occurs at a Co concentration in the vicinity of 60 % to 70 % Co. This can be observed in Fig S2b which shows a plot of the lattice constants as a function of the concentration. For samples with a Co concentration smaller than 70 %, the lattice constant for the fcc phase becomes constant at a value that is expected for pure Cu. This is in contrast to the rapidly changing value of the lattice constant observed in alloys with a concentration of Co exceeding 70 %. To put these results in context, the bulk phase diagram for this system predicts a fcc to bcc transition at approximately 75 % Co[1]. Thus, our results shows that our thin films initiate a bcc phase at a slightly higher value of Co concentration.

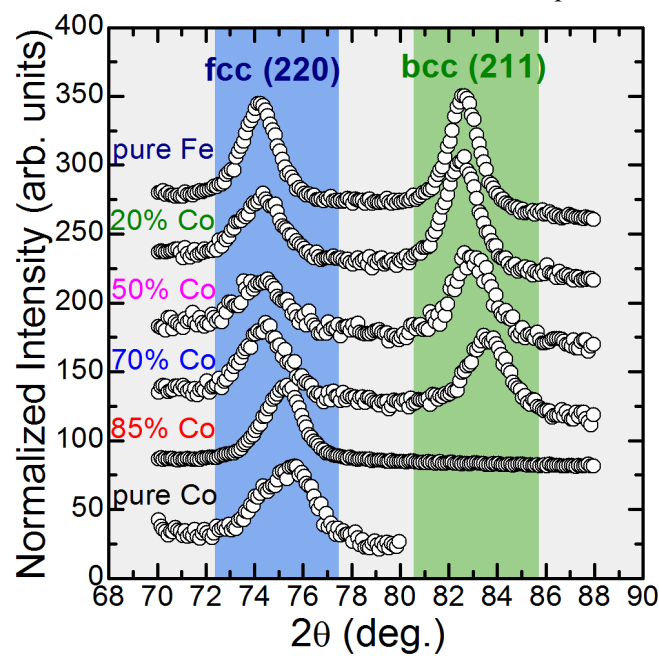

**Figure S1: XRD spectra.** The spectra are shown for select alloy concentrations. The Cu fcc peak is always visible and is overlaid with the fcc Co peak. The Co bcc peak is visible only for Co concentrations below 70%.

It is important to point out that we do not measure the expected hexagonal close-packed (hcp) crystal structure for the pure Co samples.[1] This suppression of the hcp structure was previously found in growth of Co on Cu and explained by the strained growth of the Co on Cu suppressing the hcp phase.[2–4] Indeed, we confirm this result by comparing pure Co films grown with a Ta/Cu seed and capping layer and one grown with only a Ta seed and capping layer. Figure S2a shows an XRD plot in the vicinity of the hcp(010) peak for the 10 nm pure Co samples with a Ta/Cu seed and a Ta seed. The sample with Ta seed layer exhibits a clear hcp(010) peak, indicating an hcp structure. In contrast, the sample that includes Cu in the seed layer shows no evidence of the hcp(010) peak..

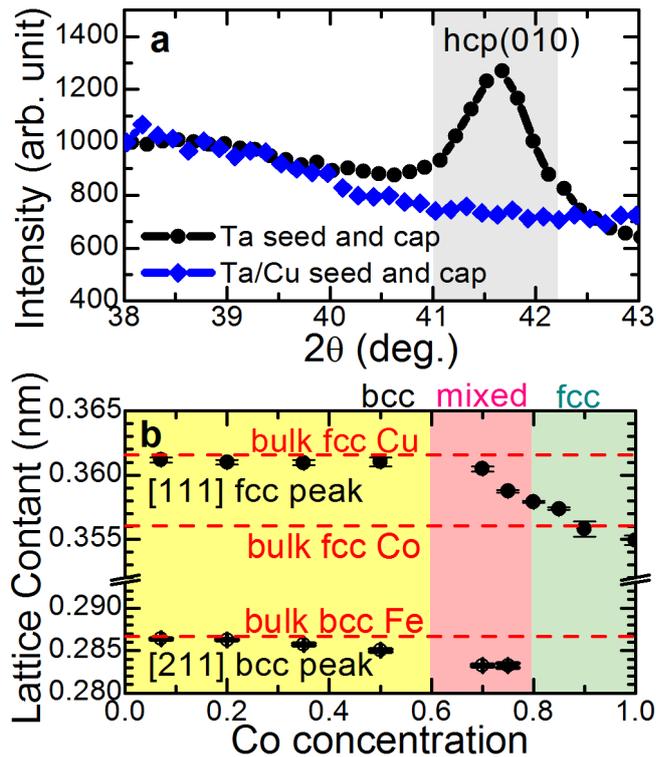

**Figure S2: XRD characterization.** In **a** the XRD spectrum of pure Co samples with a Ta and a Ta/Cu seed layer. The via XRD determined lattice constant is plotted against the respective Co concentration in **b**. The crystalline structure is denoted and coded in color.

**Saturation and effective magnetization.**

The magnetization of all of our samples were measured and verified using two approaches. The first is the direct measurement of saturation magnetization $M_s$ by superconducting quantum interference device (SQUID) magnetometry. These measured values of $M_s$ are displayed in Figure S3 as a function of the concentration of Co in the alloy. For comparison we also include the Slater-Pauling curve[1,5,6] in the plot. In the bcc phase, $M_S$ displays almost perfect agreement with the Slater-Pauling curve indicative of the high quality of the samples used in this study. However, below 80 % Co, the Slater-Pauling curve underestimates $M_s$ slightly when the crystalline phase of the CoFe alloy becomes exclusively fcc.

Ferromagnetic resonance (FMR) also provides a measurement of the effective magnetization $M_{\text{eff}}$, which is equal to $M_s - H_k$, where $H_k$ is the perpendicular anisotropy of the material. As a result of the very small thickness of our samples, we expect a non-negligible value of $H_k$ to result from the interfaces of the material with the seed and capping layers[7,8]. Indeed, $M_{\text{eff}}$ in Figure S3 are smaller than $M_s$, indicative of the presence of a perpendicular interface anisotropy. If the difference between $M_{\text{eff}}$ and $M_s$ is interfacial in origin, a thickness dependence of the $M_{\text{eff}}$ should produce the condition of $M_s = M_{\text{eff}}$ as the thickness $t \to \infty$. For a select subset of alloys, we varied the thickness of the alloy and measured

$M_{eff}$. From an extrapolation of thickness $t \to \infty$ we determined $M_s$. These values are included in Figure S3 and show reasonable agreement with the values of $M_s$ determined from SQUID magnetometry, demonstrating the equivalency of both measurement methods.

The minimum of the DOS curve at $E_F$, shown in Figure 3, is coupled to the Slater-Pauling maximum of the magnetization curve, which appears in the present work at Co concentration of approximately 25 %. For this concentration the large peak in the DOS, immediately above the Fermi level, is entirely of spin-down character, and alloys with higher Co concentration unavoidably populate this peak, which reduces the magnetization. In the present work, and in several other theoretical[9–11] and experimental[12,13] works the maximum of the magnetization (and the minimum in the density of states) appears between 20 % and 30 % Co, while in the work of Ebert[14,15] and, thus, also in the calculations of Mankovsky[16] it appears at lower Co concentrations of approximately 15 %. This may explain why in the work of Mankovsky[16] the theory predicts a minimum of the damping at this concentration.

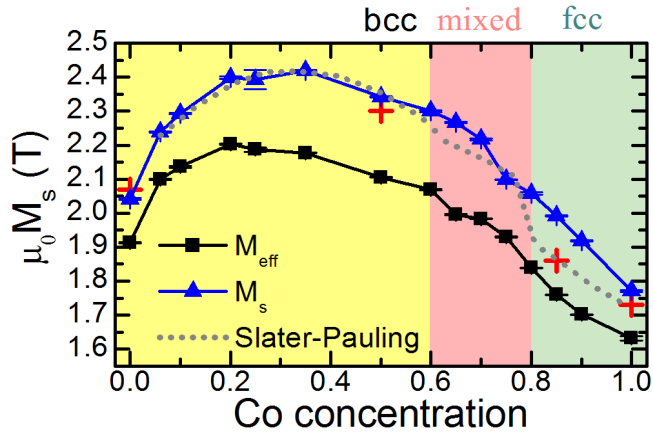

**Figure S3: The variation of the saturation magnetization and effective magnetization with concentration.** $M_{eff}$ (black squares) determined from FMR and $M_s$ (blue triangles) determined from SQUID magnetometry are plotted versus alloy concentration. For comparison, $M_s$ is also determined by extrapolating the linear regression of $M_{eff}$ vs. $1/t$, as described in the text. The extrapolated values for $M_s$ are included in the plot (red crosses) and they match $M_s$ determined by SQUID at those alloy concentrations reasonably well. This shows that SQUID and FMR measurement are consistent with each other. For comparison, the Slater-Pauling curve[1] is also shown (gray dotted line). The crystalline phase of the alloys are also indicated in the figure.

**Calculation and direct measurement of the radiative damping contribution.**

Radiative damping in the perpendicular FMR geometry, used in our study, arises from the inductive coupling of the dynamic in-plane components of the magnetization to the wave-guide. By Faraday's law, only the magnetization component perpendicular to the direction of the wave-guide can couple effectively, therefore the radiative damping is anisotropic and has to be calculated accordingly. Thus, assuming a uniform magnetization profile and excitation field in the sample, the radiative damping $\alpha_{rad}$ in our measurement is calculated via[17]

$$\alpha_{rad} = \frac{\gamma \mu_0^2 M_s t l}{8 Z_0 w_{cc}}, \qquad (S1)$$

where $\gamma = g\mu_B/\hbar$ is the gyromagnetic ratio, $Z_0 = 50\ \Omega$ the impedance, $w_{cc}$ is the width of the wave guide, and $t$ and $l$ are the thickness and length of the sample on the waveguide. This calculated value of $\alpha_{rad}$ can also be measured directly, by placing a spacer between sample and waveguide. Following the argument in Schoen et al.[17], a 100 μm glass spacer decreases $α_{rad}$ by approximately a factor of ten, making it, within errors, undetectable in our measurement. Indeed, when the spacer is inserted, the damping for both the $Co_{20}Fe_{80}$ and $Co_{25}Fe_{75}$ decreases significantly, shown in Table S1, and the damping decrease matches the calculated $α_{rad} = 0.00065$ for both samples reasonably well.

| Sample | α without spacer | α with spacer | $α_{rad}$ |
|---|---|---|---|
| $Co_{20}Fe_{80}$ | 0.00230±0.00003 | 0.0016±0.00015 | 0.0007±0.00015 |
| $Co_{25}Fe_{75}$ | 0.0021±0.00015 | 0.0018±0.0002 | 0.0003±0.00025 |

**Table S1: Direct measurement of the radiative damping contribution.** The damping of the $Co_{20}Fe_{80}$ and the $Co_{25}Fe_{75}$ sample measured with and without a 100 μm spacer between sample and waveguide. The radiative damping $α_{rad}$ is determined from the difference in damping with and without spacer.

**Determination of the interfacial damping enhancement**

The flow of spin-angular momentum into adjacent layers (in our sample geometry the Cu/Ta cap and seed layers) further enhances the measured damping. This non-local damping or spin-pumping contribution $α_{sp}$ is purely interfacial and can therefore be determined from the thickness dependence of the damping for a given alloy concentration[18]. We measured the damping of thickness series of the pure elements and select alloy concentrations (20 % Co, 25 % Co, 50 % Co and 85 % Co). In order to correctly account for the interfacial contribution $α_{sp}$ we first remove the radiative contribution $α_{rad}$ from $α_{tot}$ and in Figure S4a plot the dependence of ($α_{tot} - α_{rad}$) on the inverse thickness $1/t$ for select alloy concentrations. $α_{sp}$ is described by

$$α_{sp} = 2 \frac{g_{eff}^{↑↓} μ_B g}{4π M_S t}. \qquad (S2)$$

The factor of 2 accounts for the two nominally equal interfaces of the alloy to the seed and cap layers. The effective spin mixing conductance $g_{eff}^{↑↓}$ then is determined from the slope of the linear fits to the ($α_{tot} - α_{rad}$) vs. $1/t$ data and plotted in Figure S4b. All values of $g_{eff}^{↑↓}$ are in the range of expected values based on previous reports[19,20]. Using interpolated values of $g_{eff}^{↑↓}$ for all alloy concentrations (gray line in Figure S4b), the non-local interface effects for all alloy compositions are straightforwardly determined with equation S2.

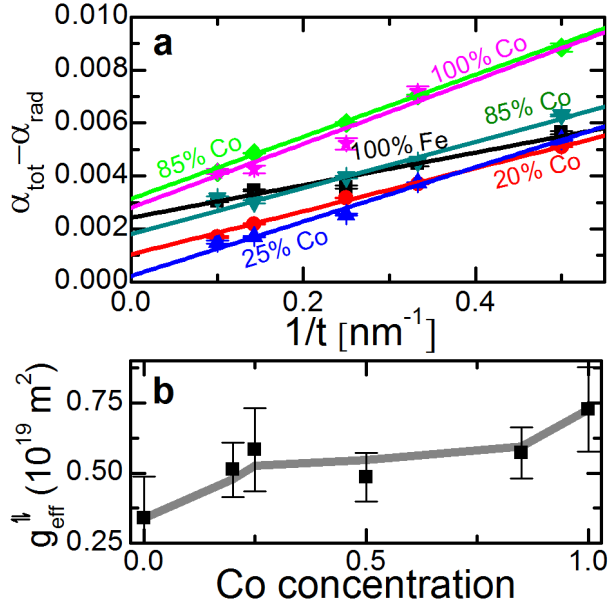

**Figure S4: Determination of the interfacial damping contribution.** The ($\alpha_{tot}$ - $\alpha_{rad}$) vs. $1/t$ dependence (symbols) for the pure elements and select alloy concentrations (20% Co, 25% Co, 50% Co and 85% Co) with fits to Equation 2 (lines) is plotted in **a**. From these fits the effective spin mixing conductance $g_{eff}^{\uparrow\downarrow}$ is calculated and plotted in **b**. The gray line is an interpolation of the calculated values.

## Comparison of $\alpha_{int}$ to other theories

Gilmore et al.[21] calculated the damping as a function of the electron-phonon self-energy $\Gamma$. They found a minimum in damping for the pure elements bcc Fe ($\alpha_{int}$ = 0.0013), hcp Co ($\alpha_{int}$ = 0.0011) and fcc Ni ($\alpha_{int}$ = 0.017). A comparison of these values to the extracted $\alpha_{int}$ for bcc Fe $\alpha_{int}$ = 0.0024) and fcc Co ($\alpha_{int}$ = 0.0026) shows that these calculated values are well below what we estimate to be the intrinsic damping. However, an adjustment of the electron-phonon self-energies $\Gamma$ to 3 meV for Fe and 2 meV for Co covers our findings. Nevertheless, we acknowledge that our sputtered films contain some disorder and strain. Thus, it is conceivable that epitaxially grown, single crystal CoFe alloys may exhibit even lower values of damping.